\documentclass[prb,twocolumn,showpacs,floatfix]{revtex4}
\usepackage{graphicx}
\usepackage{dcolumn}
\usepackage{bm}
\begin{document}
\title{Direct evidence for the magnetic ordering of Nd ions in NdFeAsO by high resolution inelastic neutron scattering}
\author{T. Chatterji$^1$, G. N. Iles$^1$, B. Frick$^1$, A. Marcinkova$^2$ and J.-W. G. Bos$^3$}
\address{$^1$Institut Laue-Langevin, 6 rue Joules Horowitz, BP 156, 38042 Grenoble Cedex 9, France\\
$^2$School of Chemistry and Centre for Science at Extreme Conditions, University of Edinburgh, West Mains Road, Edinburg EH9 3 JJ, United Kingdom\\
$^3$Centre for Advanced Energy Storage and Recovery, School of Engineering and Physical Sciences - Department of Chemistry, Heriot-Watt University, Edinburgh EH14 4AS, United Kingdom.}
\date{\today}

\begin{abstract}
We investigated the low energy excitations in the parent compound NdFeAsO of the Fe-pnictide superconductor 
in the $\mu$eV range by a back scattering neutron spectrometer. The 
energy scans on a  powder NdFeAsO sample revealed inelastic peaks at 
E = 1.600 $ \pm 0.003 \mu$eV  at T = 0.055 K  on both energy gain and 
energy loss sides. The 
inelastic peaks move gradually towards lower energy with increasing 
temperature and finally merge with the elastic peak  at about $6$ K. 
We interpret the inelastic peaks to be due to the transition between 
hyperfine-split nuclear level  of the $^{143}$Nd and $^{145}$Nd isotopes 
with spin $I = 7/2$. The hyperfine field is produced by the ordering of the electronic magnetic moment of Nd at low temperature and thus the present investigation gives direct evidence of the ordering of the Nd magnetic sublattice of NdFeAsO at low temperature.
\end{abstract}
\maketitle
    The recent discovery of superconductivity in doped iron-arsenide-oxides \cite{kamikahara08} has heralded a new era in superconductivity research \cite{johrendt08}. After the first report that LaFeAsO$_{1-x}$F$_x$ became a superconductor below $T_c = 26$ K, related compounds with higher critical temperatures, up to 55 K in SmFeAsO$_{1-x}$F$_x$, were quickly developed \cite{ren08}. It is now accepted that these materials comprise a second class of high-$T_c$ superconductors, distinct from the superconducting cuprates discovered more than 20 years ago \cite{bednorz86}. The parent compound LaFeAsO crystallizes in the tetragonal ZrCuSiAs-type structure (Space group $P4/nmm$) \cite{johnson74}. It is currently believed that the superconductivity in doped LaFeAsO is, like in cuprates, is not conventional BCS-type but intimately connected with magnetic fluctuations and with a spin density wave (SDW) anomaly within the FeAs layers \cite{dong08,chen08}. Undoped LaFeAsO undergoes a SDW-driven structural phase transition around 150 K, in which the lattice symmetry drops from tetragonal to orthorhombic \cite{nomura08} and anomalies in specific heat, electrical resistance and magnetic susceptibility occur. The antiferromagnetic ordering of the magnetic moments ($0.36 \mu_B$/Fe) was found below $T_N = 134$ K by neutron scattering \cite{cruz08}. Electron doping by F or by O deficiency, as well as hole doping with Sr suppresses the phase transition and the tetragonal phase becomes superconducting at $T_c $ in the temperature range 25-41 K \cite{kamikahara08,lu08,wen08}. Thus there is strong evidence that superconductivity in LaFeAsO is mainly due to specific structural and electronic properties of the (FeAs)$^{\delta-}$ layers.The magnetic structures of the RFeAsO family have been investigated by neutron diffraction. As we mentioned before the Fe moments order below about 150 K accompanied by a tetragonal-to-orthorhombic phase transition at about the same temperature. The rare-earth moments order at lower temperatures. The magnetic ordering of NdFeAsO has been investigated by specific heat \cite{baker09} and neutron diffraction \cite{qiu08,chen08a,tian10,marcinkova10}. NdFeAsO undergoes \cite{tian10} a tetragonal-to-orthorhombic structural phase transition at $T_S \approx 142$ K. The Fe moments order antiferromagnetically in the stripe-like structure in the ($a$-$b$) plane below $T_N = 137$ K but change from antiferromagnetic (AFM) to the ferromagnetic (FM) arrangement along the $c$ axis below $T_*\approx 15$ K.  The Nd moments order antiferromagnetically below about $6$ K with the same structure. However neutron diffraction measurements suggest that the Nd moments are polarised below $T^*\approx 15$ K.  Here we report a high resolution inelastic neutron scattering investigation of the Nd magnetic ordering by studying the transition between hyperfine split nuclear levels of Nd nucleus generated by the hyperfine field at the Nd nucleus due to the ordered electronic magnetic moment of Nd ions. This relatively less-known technique has been recently developed and tested in several Nd-based compounds \cite{chatterji00,chatterji02,chatterji04,chatterji04a,chatterji08,chatterji08a,chatterji09}. The inelastic peaks observed in NdFeAsO below about $T_{Nd} \approx 6$ K show directly the magnetic ordering of Nd ions at low temperature. From these studies we established that the energy of nuclear spin excitations is proportional to the ordered magnetic moment in Nd compounds. Here we used this linear relationship to estimate the ordered magnetic moment of Nd ions in NdFeAsO.  By this technique we only observe the inelastic signal due to the magnetic ordering of Nd ions because the spin-dependent neutron scattering cross section of $^{57}$Fe is too small ($0.5\pm 0.1$ barn for  $^{57}$Fe compared to $56 \pm 3$ barn for $^{143}$Nd ) to observe the signal due to the ordering of Fe moments. So the presence of inelastic signals in NdFeAsO at low temperature gives direct evidence for the magnetic ordering of Nd ions. 
    
    The principle of the method of studying the hyperfine interaction by the inelastic neutron scattering
can be summarized \cite{heidemann70,heidemann72} as follows: If  neutrons with spin ${\bf s}$ 
are scattered from  nuclei with spins  ${\bf I}$, the probability that 
their spins will be flipped is $2/3$. The nucleus at which the 
neutron is scattered with a spin-flip, changes its magnetic quantum 
number $M$ to $M\pm 1$ due to the conservation of the 
angular momentum. If the nuclear ground state is split up into 
different energy levels $E_{M}$ due to the 
hyperfine magnetic field or an electric quadrupole interaction, then 
the neutron spin-flip produces a change of the ground state energy 
$\Delta E = E_{M} - E_{M\pm 1}$. This energy change is transferred 
to the scattered neutron. The double differential scattering cross section \cite{heidemann70} is 
given by  the following expressions:
\begin{equation}
	 \left(\frac{d^2\sigma}{d\Omega d\omega}\right)_
{inc}^{0}=\overline{(\overline{\alpha^{2}}-{\overline{\alpha}}^{2}+
\frac{1}{3}{\alpha^{\prime}}^{2}I(I+1))}e^{-2W(Q)}
\delta(\hbar\omega),
\label{heidemann01}
\end{equation}
\begin{equation}
\left(\frac{d^2\sigma}{d\Omega d\omega}\right)_
{inc}^{\pm}=
\frac{1}{3}\overline{{\alpha^{\prime}}^{2}I(I+1)}\sqrt{1\pm\frac{\Delta E}{E_{0}}}e^{-2W(Q)}
\delta(\hbar\omega\mp \Delta E)
\label{heidemann02}	 
\end{equation}
where $\alpha$ and $\alpha^{\prime}$ are coherent and spin-incoherent 
scattering lengths, $W(Q)$ is the Debye-Waller factor and $E_{0}$ is 
the incident neutron energy, $\delta$ is the Dirac delta function.  
The long bar signs on top of equation (\ref{heidemann01}) and (\ref{heidemann02}) mean averages of the quantity on top of which they stay. If the sample contains one type of isotope then 
$\overline{\alpha^{2}}-{\overline{\alpha}}^{2}$ is zero. Also 
$\sqrt{1\pm\frac{\Delta E}{E_{0}}} \approx 1$ because $\Delta E$ is 
usually much less than the incident neutron energy $E_{0}$. In this case 2/3 of 
incoherent scattering will be spin-flip scattering. The 
hyperfine splitting lies typically in the energy range of a few $\mu 
eV$. The inelastic spin-flip scattering of neutrons from the nuclear spins can 
yield this information provided the neutron spectrometer has the 
required resolution of about $1 \mu eV$ or less and also the incoherent 
scattering of the nucleus is strong enough. Because of relatively large spin incoherent cross section of natural Nd, the Nd-based compounds are 
very much suitable for the studies of nuclear spin excitations. The Nd has the natural abundances of 12.18\% 
and 8.29\% of $^{143}$Nd and $^{145}$Nd isotopes, respectively. Both 
of these isotopes have nuclear spin of $I = 7/2$ and their incoherent 
scattering cross sections \cite{sears99} are relatively large, $55\pm 7$ and $5\pm 5$ barn
for $^{143}$Nd and $^{145}$Nd , respectively. Taking into account the natural abundance of these isotopes the total spin incoherent scattering of the natural Nd is $\sigma _i$(spin)$ = 7.1$ barn.
\begin{figure}
\resizebox{0.5\textwidth}{!}{\includegraphics{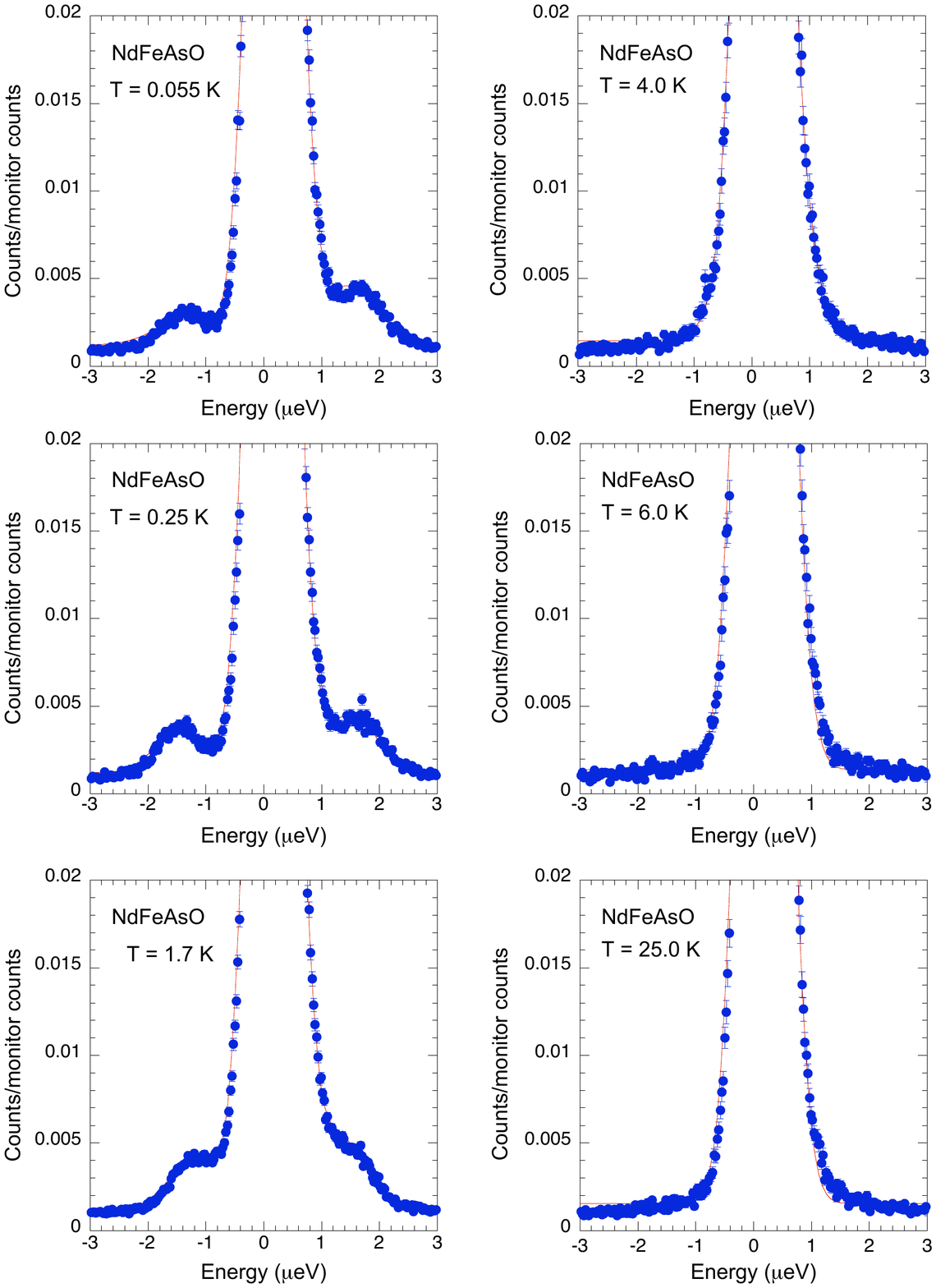}}

\caption { Energy spectra of NdFeAsO at several temperatures. 
           } 
 \label{spectra}
\end{figure}

\begin{figure}
\resizebox{0.5\textwidth}{!}{\includegraphics{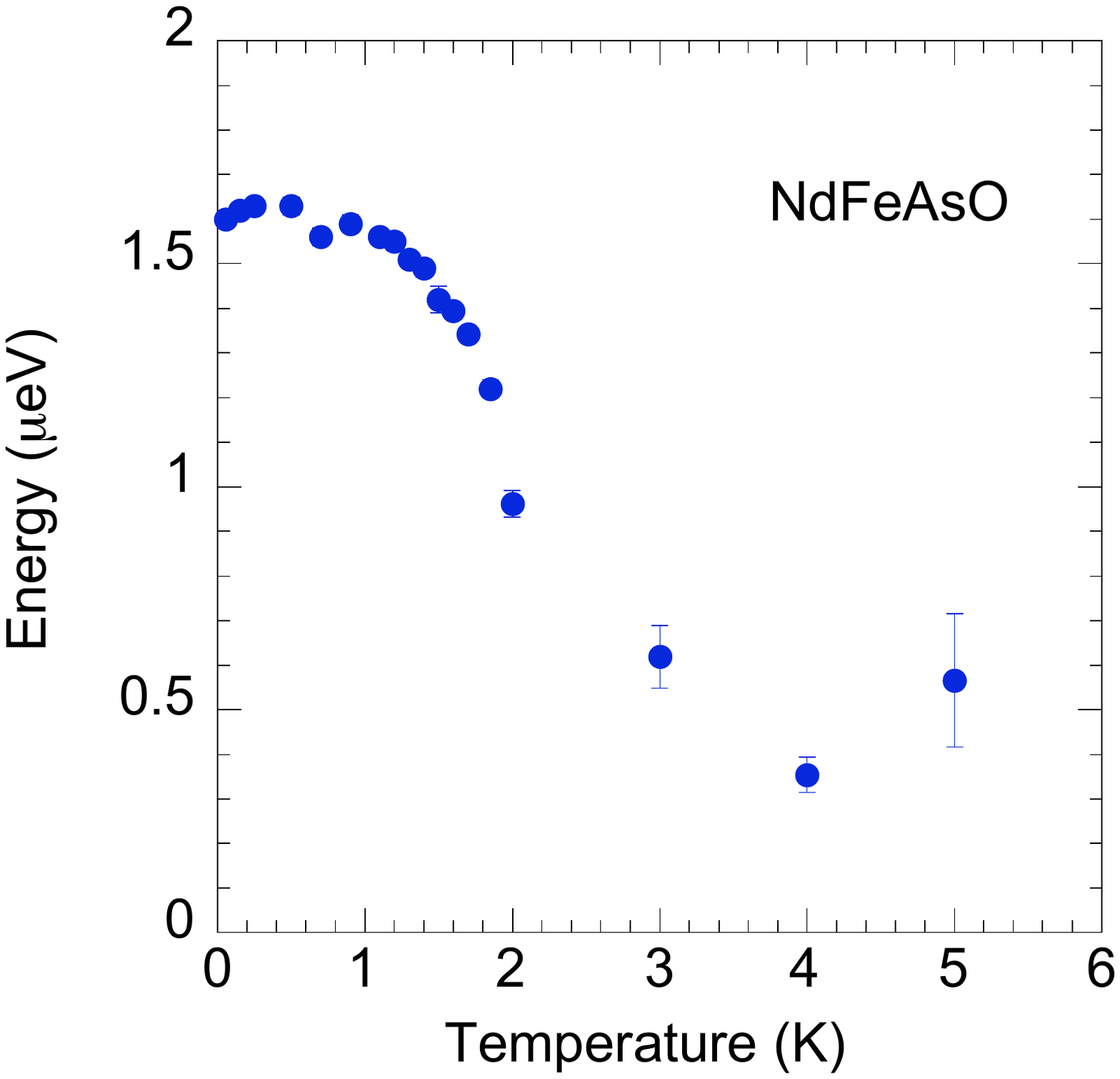}}
\caption {Temperature variation of the energy of the inelastic peak 
          of NdFeAsO.  } 
\label{Tdependence}
\end{figure}

\begin{figure}
\resizebox{0.5\textwidth}{!}{\includegraphics{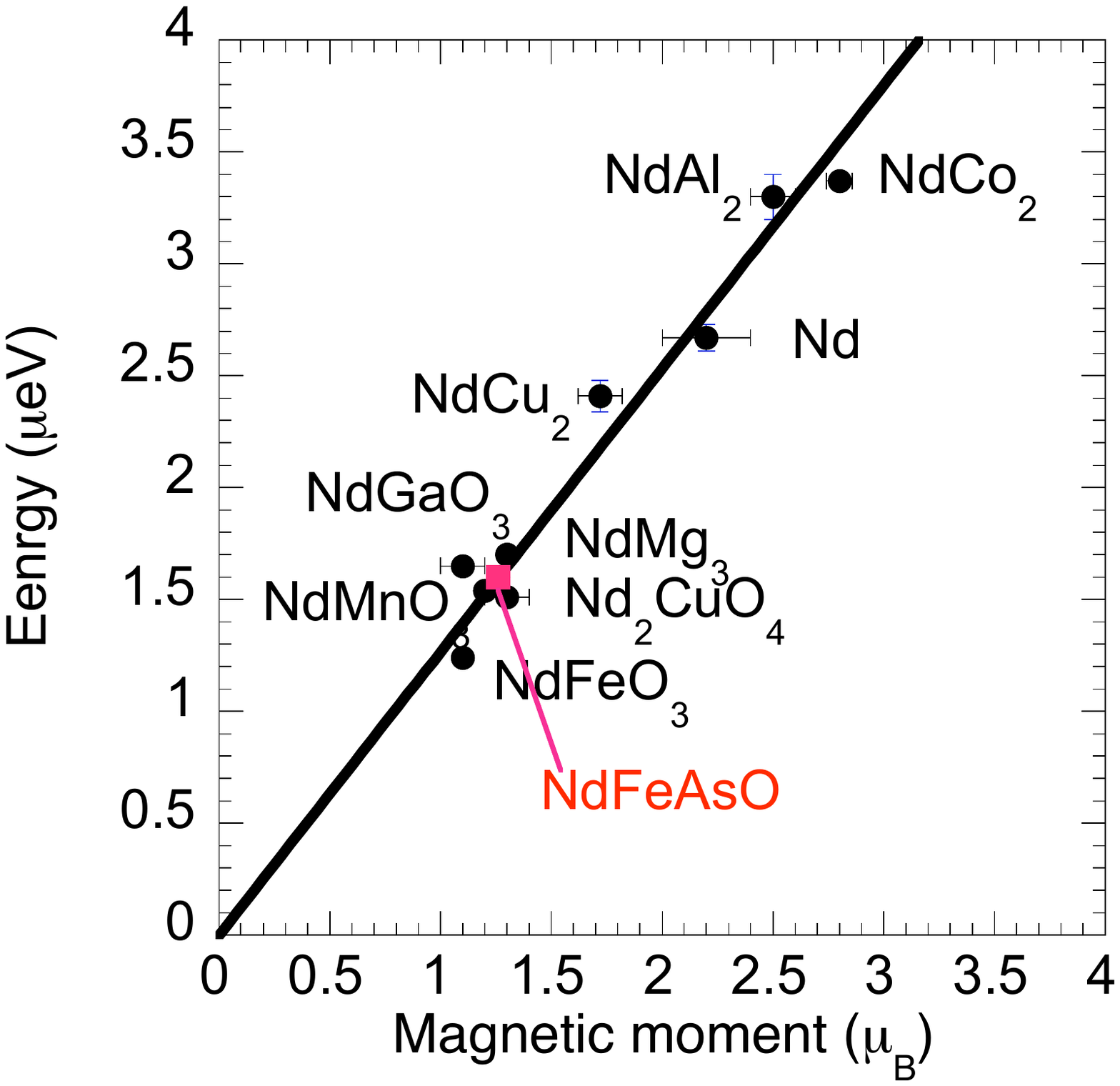}}
 \caption { Plot of energy of the inelastic signals in Nd$_2$CuO$_{4}$ \cite{chatterji00}, Nd 
metal \cite{chatterji02}, NdCu$_{2}$ \cite{chatterji04}, NdGaO$_{3}$ \cite{chatterji04a},  NdFeO$_3$ \cite{przenioslo06}, NdMg$_3$ \cite{chatterji08}, NdCo$_2$ \cite{chatterji08},
NdMnO$_3$ \cite{chatterji08a}  and NdAl$_2$ \cite{chatterji09} vs. the corresponding 
electronic magnetic moment of Nd in these compounds determined by 
the refinement of the magnetic structure using magnetic neutron diffraction
intensities. We used the energy of the nuclear spin excitation determined in the present experiment to determine the magnetic moment of Nd ion in NdFeAsO at $T = 0.055$ K by using the linear plot. The red filled square in the plot shows the corresponding data for NdFeAsO. The magnetic moment determined from the linear plot is $1.26 \mu_B$.} 
\label{ndmoment}
\end{figure}

A Polycrystalline NdFeAsO sample was prepared as described before \cite{marcinkova09}. The lattice constants (a = 3.966, c = 8.589 {\AA}, V = 135.1 {\AA$^3$}) were extracted from Rietveld fits against laboratory powder X-ray diffraction data and are in excellent agreement with our previous work \cite{marcinkova09,marcinkova10} and recently reported data on crushed single crystals \cite{yan11}. These fits also revealed minor Nd$_2$O$_3$ and FeAs impurities with a total weight fraction less than 5\%, which is comparable to Ref 15, and which evidently do not affect the lattice dimensions of the main phase. No evidence for the presence of Fe was found. About 5 g of powder NdFeAsO sample thus prepared was placed inside a flat Al sample holder and was attached to the sample stick  of a standard $^4$He cryostat of the high-resolution back-scattering neutron spectrometer IN16 of the Institut Laue-Langevin in Grenoble. The neutron wavelength was $\lambda = 6.27$ {\AA}. The inelastic signals from the sample were measured at several temperatures in the temperature range $1.7 - 25$ K. In order to measure the inelastic signal from NdFeAsO at lower temperature down to about $55$ mK the sample was then placed inside the annular space of a Cu double-walled cylindrical sample holder and was attached to the $^3$He-$^4$He dilution insert of the cryostat. 

Fig. \ref{spectra} shows the inelastic spectra from NdFeAsO sample at several temperatures.  The data for all Q has been integrated. We however checked carefully the Q dependence of the signals and  convinced ourselves hat no such dependence exists. The measured Q range was from about 0.2 to 1.9 {\AA}$^{-1}$. The lack of appreciable Q dependence of the inelastic signals due hyperfine interaction is expected and is checked by us in a variety of Nd-based compounds. At $T = 0.055$ K clear inelastic signals were seen at $E = 1.600 \pm 0.003$ $\mu$eV on both energy loss and energy gain sides. The inelastic signals move to the central elastic peak at higher temperature and finally almost merg with it at $T_{Nd} \approx 6$ K. However the inelastic signals at a very small energy seem to persist at higher temperature but could not be evaluated due to their proximity to the central elastic peak. At $T= 25$ K the signal is purely elastic and its width is explained by the resolution function of the spectrometer. The spectra have been fitted to the convoluted resolution function of the spectrometer measured by the signal from the NdFeAsO sample at $T = 25$ K.  Fig. \ref{Tdependence} shows the temperature dependence of the energy of nuclear excitations. The fitting procedure does not work any more at $T = 6$ K due to the proximity of the signal to the central elastic peak as we already noted before. The temperature variation of the energy of nuclear spin excitations shows that the Nd ordering process consists of two steps. At temperatures above 2 K the Nd ordering is perhaps induced by the increasing Nd-Fe exchange interaction whereas below 2 K the ordering is due to the onset of the Nd-Nd interaction. Unfortunately due to the limited neutron beam time we could not measure the spectra at smaller temperature intervals which would establish the two stages of the Nd ordering process more clearly.
At millikelvin temperature the effect of detailed balance is observed in the inelastic signals in the energy loss and energy gain sides. If one compares carefully the spectra at $T = 0.055$ K and $T = 0.25$ K in Fig. \ref{spectra} then one will notice that at $T = 0.055$ K the inelastic signal in the energy loss side is weaker than that in the energy gain side whereas at $T = 0.25$ K the signals on both sides are about equal.

We have studied hyperfine interactions  in several Nd-based compound recently by the high resolution neutron scattering technique and have found a linear relationship between the energy of nuclear spin excitations at low temperature and the ordered magnetic moment of the Nd ions in these compounds. One expects at low temperature in insulating Nd compounds an ordered magnetic moment of $3 \mu_B$ but due to the crystal-field effects one obtains a much lower ordered magnetic moments in neutron diffraction experiments. In Figure \ref{ndmoment} we show a plot of energy of inelastic peaks in Nd$_2$CuO$_{4}$ \cite{chatterji00}, Nd 
metal \cite{chatterji02}, NdCu$_{2}$ \cite{chatterji04}, NdGaO$_{3}$ \cite{chatterji04a},  NdFeO$_3$ \cite{przenioslo06}, NdMg$_3$ \cite{chatterji08}, NdCo$_2$ \cite{chatterji08}
NdMnO$_3$ \cite{chatterji08a}  and NdAl$_2$ \cite{chatterji09} vs. the corresponding 
electronic magnetic moment of Nd ions in these compounds determined by 
the refinement of the magnetic structure using magnetic neutron diffraction
intensities. The data lie approximately on a straight line 
showing that the hyperfine field at the nucleus is approximately 
proportional to the electronic magnetic moment. The slope of the linear fit of the
data gives a value of $1.27 \pm 0.03 \mu eV$/$\mu_B$. It is to be noted that the data 
for the hyperfine splitting is rather accurate whereas the magnetic 
moments determined by neutron diffraction have large standard 
deviations and are dependent on the magnetic structure models. The magnetic 
structures are seldom determined unambiguously and the magnetic moment determined
from the refinement of a magnetic structure model is  relatively uncertain. In such cases the investigation of the low energy excitations
described here can be of additional help \cite{chatterji08a}. This is specially true for the complex magnetic structures
with two magnetic sub-lattices of which one sublattice contains Nd. Such complex magnetic  
structures, such as the parent compounds of newly discovered Fe-based superconductors, colossal magnetoresistive manganites and some multiferroic materials, are currently under intense
study. Here we have used the nuclear spin excitation energy of NdFeAsO determined from the present experiment to determine the magnetic moment of Nd ion in NdFeAsO at $T = 0.055$ K by using the linear plot shown in Fig. \ref{ndmoment}. The magnetic moment determined from the linear plot is $1.26 \mu_B$. Thus the magnetic moment of Nd ions are reduced considerably by the crystal field effect. The ordered magnetic moment of Nd in NdFeAsO has not yet been unambigusly reported by neutron diffraction. The present estimation of the ordered magnetic moment of Nd in NdFeAsO remains to be checked by neutron diffraction experiment at very low temperature. Detailed neutron diffraction investigation on NdFeAsO single crystals is urgently needed. It would also be useful to study the crystal field effects in NdFeAsO by inelastic neutron scattering.

In conclusion we have observed inelastic neutron scattering signals at low temperature in NdFeAsO, which give direct evidence for the magnetic ordering of Nd ions and we have estimated from the linear relationship established for Nd-based compounds the ordered magnetic moment of Nd  to be  $1.26 \mu_B$. This estimate of the Nd magnetic moment is independent of the magnetic structure model.

\end{document}